\documentclass[english]{article}
\usepackage{amssymb}

\usepackage{fontenc}
\usepackage[latin1]{inputenc}
\usepackage{amsmath}
\usepackage{fontenc}
\usepackage{babel}

\input{tcilatex}

\begin{document}

\title{\textbf{Energy-momentum tensor in thermal strong-field QED with unstable
vacuum}}
\author{S.P. Gavrilov\thanks{%
Electronic address: gavrilovsergeyp@yahoo.com} and D.M. Gitman \thanks{%
Instituto de Física, Universidade de São Paulo, CP 66318, CEP 05315-970 São
Paulo, SP, Brazil; E-mail: gitman@dfn.if.usp.br} \\
Department of General and Experimental Physics,\\
Herzen State Pedagogical University of Russia, \\
Moyka emb. 48, 191186 St. Petersburg, Russia}
\maketitle

\begin{abstract}
The mean value of the one-loop energy-momentum tensor in thermal QED with an
electric-like background that creates particles from vacuum is calculated.
The problem is essentially different from calculations of effective actions
(similar to the action of Heisenberg--Euler) in backgrounds that respect the
stability of vacuum. The role of a constant electric background in the
violation of both the stability of vacuum and the thermal character of
particle distribution is investigated. Restrictions on the electric field
and its duration under which one can neglect the back-reaction of created
particles are established.

\end{abstract}

I. Quantum field theory (QFT) in external backgrounds is a good model for
the study of quantum processes in cases when a part of a quantized field is
strong enough to be treated as a classical one. The validity of that
approach is based on the assumption that back-reaction is small. However,
from physical reasons, it is clear that back-reaction may be quite strong in
external backgrounds that can create particles from vacuum and produce
actual work on particles. A complete description of back-reaction is
connected to the calculation of the mean energy-momentum tensor (EMT) for
the matter field. In this connection, we recall that Heisenberg and Euler
computed the mean value of energy density of the Dirac field for a constant
weak ($\left| E\right| \ll E_{c}=M^{2}c^{3}/|e|\hbar \simeq 1,3\cdot
10^{16}\;V/cm$) electromagnetic background at zero temperature and presented
the respective one-loop effective Lagrangian $\mathcal{L};$ see \cite{HeiE36}%
. This Lagrangian was re-introduced by Schwinger \cite{S51}, without any
restrictions on the external field, in order to represent the
vacuum-to-vacuum transition amplitude, $c_{v}$, as well as the probability, $%
P^{v}$, of a vacuum to remain a vacuum in a constant electric field, namely, 
\begin{equation}
c_{v}=\exp \left( i\int dx\mathcal{L}\right) ,\;P^{v}=|c_{v}|^{2}=\exp \{-VT2%
\func{Im}\mathcal{L}\}\,.  \label{vacampl}
\end{equation}
Here, $T$ is the field duration, and $V$ is the observation volume.
Nevertheless, the problem of back-reaction can be solved only by computing
the mean EMT in the corresponding background. This kind of calculations
first appeared in \cite{Gav05} for a spinor field in arbitrary constant
electric background with the vacuum initial state, and then developed and
completed in \cite{GavG07}, including the case of thermal initial state.

Here, on the basis of these results, we discuss the mean EMT $\langle T_{\mu
\nu }\left( t\right) \rangle $ of a spinor field in a quasiconstant electric
background. Namely, we select the background as a so-called $T$-constant
field, being a uniform constant electric field $E$, acting only during a
sufficiently large but finite time interval $T$, 
\begin{equation}
T>>T_{0},\;T_{0}=\left| qE\right| ^{-1/2}+M^{2}\left| qE\right| ^{-3/2},
\label{unistab}
\end{equation}
where $q$ is the charge and $M$ is the mass of an electron. The\ $T$%
-constant field turns on at $t_{1}=-T/2;$ it turns off at $t_{2}=T/2$, and
is directed along the $x^{3}$-axis\footnote{%
As shown in \cite{GavG96a}, calculations of particle creation in a $T$%
-constant field explains typical features of the effect in a large class of
quasiconstant electric backgrounds.}. In addition, a parallel constant
magnetic field $B$ may also be present at the background. We treat the
background nonperturbatively and use the technics developed in \cite
{FGS91,FraGi81} and \cite{GavGF87,GavGT06}.

II. The mean EMT $\langle T_{\mu \nu }\left( t\right) \rangle $ is defined
as 
\begin{equation}
\langle T_{\mu \nu }\left( t\right) \rangle =\mathrm{Tr\,}\left[ \rho
_{in}T_{\mu \nu }\right] \,,  \label{emt22}
\end{equation}
where $\rho _{in}$ is the density operator of the initial state in the
Heisenberg picture; the EMT operator $T_{\mu \nu }$ has the form 
\begin{equation*}
T_{\mu \nu }=\frac{1}{8}\left\{ \left[ \bar{\psi}(x),\gamma _{\mu }P_{\nu
}\;\psi (x)\right] +\left[ P_{\nu }^{\ast }\bar{\psi}(x),\gamma _{\mu
}\;\psi (x)\right] \right\} \,+\frac{1}{8}\left\{ \mu \leftrightarrows \nu
\right\} ,
\end{equation*}
$P_{\mu }=i\partial _{\mu }-qA_{\mu }(x),$ while $\psi (x)$ denote spinor
field operators (in the Heisenberg representation) which obey the Dirac
equation with the corresponding external field. The initial state of the
system under consideration is selected as an equilibrium state of
noninteracting $in$-particles at the temperature $\theta $ with the chemical
potentials $\mu ^{(\zeta )},$ so that 
\begin{equation}
\rho _{in}=Z^{-1}\exp \left\{ \beta \left( \sum_{\zeta =\pm }\mu ^{(\zeta
)}N^{(\zeta )}-H\left( t_{1}\right) \right) \right\} ,\;\;\mathrm{Tr\,}\rho
_{in}=1\,,  \label{emt1}
\end{equation}
where $\beta =\theta ^{-1},$ and $N^{(\zeta )}$ are the $in$-paricle number
operators: 
\begin{equation*}
N^{(+)}=\sum_{n}a_{n}^{\dagger }\left( in\right) a_{n}\left( in\right)
\,,\;N^{(-)}=\sum_{n}b_{n}^{\dagger }\left( in\right) b_{n}\left( in\right)
\,.
\end{equation*}
It is supposed that in the Heisenberg picture (see \cite{FGS91,GavGT06} for
notation) there exists a set of creation and annihilation operators, $%
a_{n}^{\dagger }(in)$, $a_{n}(in)$ of $in$-particles (electrons), and
similar operators $b_{n}^{\dagger }(in)$, $b_{n}(in)$ of $in$-antiparticles
(positrons), with a corresponding $in$-vacuum $|0,in\rangle ,$ and a set of
creation and annihilation operators $a_{n}^{\dagger }(out)$, $a_{n}(out),$
of $out$-electrons, and similar operators $b_{n}^{\dagger }(out)$, $%
b_{n}(out)$ of $out$-positrons, with a corresponding $out$-vacuum $%
|0,out\rangle $. By $n$ we denote a complete set of possible quantum
numbers. The $in$- and $out$-operators obey the canonical anticommutation
relations. The Hamiltonian $H\left( t\right) $ of the quantized spinor
field\ is time-dependent due to the external field. It is diagonalized (and
also has a canonical form) in terms of the first set at the initial instant
of time, and is diagonalized (and has a canonical form) in terms of the
second set at the final instant of time.

For example, the quantum Heisenberg field $\psi (x)$ can be expressed in
terms of the creation and annihilation operators of $in$-particles with the
help of some appropriate sets of solutions of the Dirac equation with the
external field. Namely, the $in$-particles are associated with a complete
set ($in$-set) of solutions $\left\{ _{\zeta }\psi _{n}(x)\right\} $ with
asymptotics $_{\zeta }\psi _{n}(t_{1},\mathbf{x})$ (at the initial time
instant $t_{1}$) being eigenvectors of the one-particle Dirac Hamiltonian $%
\mathcal{H}(t)=\gamma ^{0}(\left[ M+\boldsymbol{\gamma }\left( i%
\boldsymbol{\nabla }-q\mathbf{A}\left( t,\mathbf{x}\right) \right) \right] )$
at $t=t_{1}$, 
\begin{equation}
\mathcal{H}(t_{1})_{\zeta }\psi _{n}(t_{1},\mathbf{x})=\zeta \varepsilon
_{n}^{(\zeta )}{}_{\zeta }\psi _{n}(t_{1},\mathbf{x})\,,  \label{asy1}
\end{equation}
where $\varepsilon _{n}^{(\zeta )}$\ are the energies of $in$-particles in a
state specified by a complete set of quantum numbers $n$, and $\varepsilon
_{n}^{(\pm )}>0$. Then, 
\begin{equation}
\psi (x)=\sum_{n}{}\left[ _{+}\psi _{n}(x)a_{n}(in)+{}_{-}\psi
_{n}(x)b_{n}^{\dagger }(in)\,\right] {}  \label{emt2a}
\end{equation}
and 
\begin{equation}
H\left( t_{1}\right) =\sum_{n}\left[ \varepsilon _{n}^{(+)}a_{n}^{\dagger
}\left( in\right) a_{n}\left( in\right) +\varepsilon
_{n}^{(-)}b_{n}^{\dagger }\left( in\right) b_{n}\left( in\right) \right] \,.
\label{emt2}
\end{equation}
Correspondingly, the initial vacuum is defined by $a_{n}(in)|0,in\rangle
=b_{n}(in)|0,in\rangle =0$ for every $n$.

Using representation (\ref{emt2a}), we can express the Green function $i%
\mathrm{Tr\,}\left\{ \rho _{in}T\psi \left( x\right) \bar{\psi}\left(
x^{\prime }\right) \right\} $ via the $in$-set of solutions, separating at
the same time the temperature-dependent contribution from the vacuum
contribution as follows: 
\begin{equation}
i\mathrm{Tr\,}\left\{ \rho _{in}T\psi \left( x\right) \bar{\psi}\left(
x^{\prime }\right) \right\} =S^{\theta }\left( x,x^{\prime }\right)
+S_{in}^{c}\left( x,x^{\prime }\right) \,,  \label{et1}
\end{equation}
where 
\begin{eqnarray}
&&S_{in}^{c}(x,x^{\prime })=i\langle 0,in|T\psi (x)\bar{\psi}(x^{\prime
})|0,in\rangle \,  \notag \\
&&\,=\theta \left( x_{0}-x_{0}^{\prime }\right) S_{in}^{-}\left( x,x^{\prime
}\right) -\theta \left( x_{0}^{\prime }-x_{0}\right) S_{in}^{+}\left(
x,x^{\prime }\right) \,,  \notag \\
&&S_{in}^{\mp }\left( x,x^{\prime }\right) =i\sum_{n}\,_{\pm }\psi
_{n}\left( x\right) _{\pm }\bar{\psi}_{n}\left( x^{\prime }\right) \,;
\label{et2}
\end{eqnarray}
and 
\begin{eqnarray}
S^{\theta }\left( x,x^{\prime }\right) &=&i\sum_{n}\left[ -\ _{+}\psi
_{n}\left( x\right) _{+}\bar{\psi}_{n}\left( x^{\prime }\right)
N_{n}^{(+)}\left( in\right) +\ _{-}\psi _{n}\left( x\right) _{-}\bar{\psi}%
_{n}\left( x^{\prime }\right) N_{n}^{(-)}\left( in\right) \right] ,  \notag
\\
N_{n}^{\left( \zeta \right) }\left( in\right) &=&\left[ \exp \left\{ \beta
\left( \varepsilon _{n}^{(\zeta )}-\mu ^{(\zeta )}\right) \right\} +1\right]
^{-1}\,.  \label{et3}
\end{eqnarray}
It is important to stress that due to vacuum instability $S_{in}^{c}$
differs from the causal Green function 
\begin{equation}
S^{c}(x,x^{\prime })=i\frac{\langle 0,out|T\psi (x)\bar{\psi}(x^{\prime
})|0,in\rangle }{\langle 0,out|0,in\rangle }\,.  \label{emt17}
\end{equation}
This causal Green function and the difference $S^{p}(x,x^{\prime })=\
S_{in}^{c}(x,x^{\prime })-S^{c}(x,x^{\prime })$\ can be expressed through
appropriate solutions of the Dirac equation. To this end, we define that the 
$out$-particles be associated with a complete set ($out$-set) of solutions $%
\left\{ ^{\zeta }\psi _{n}(x)\right\} $ with asymptotics $^{\zeta }\psi
_{n}(t_{2},\mathbf{x})$ (at the final time instant $t_{2}$) being
eigenvectors of the one-particle Dirac Hamiltonian $\mathcal{H}(t_{2})$. The 
$out$-set can be decomposed in the $in$-set as follows: 
\begin{equation}
{}^{\zeta }\psi (x)={}_{+}\psi (x)G\left( {}_{+}|{}^{\zeta }\right)
+{}_{-}\psi (x)G\left( {}_{-}|{}^{\zeta }\right) \,,  \label{emt18}
\end{equation}
where the coefficients $G\left( {}_{\zeta }|{}^{\zeta ^{\prime }}\right) $
are expressed via inner products of these sets. These coefficients obey
unitary conditions that follow from normalization conditions imposed on the
solutions. From (\ref{emt17}), it follows that 
\begin{eqnarray}
&&S^{c}\left( x,x^{\prime }\right) =\theta \left( x_{0}-x_{0}^{\prime
}\right) S^{-}\left( x,x^{\prime }\right) -\theta \left( x_{0}^{\prime
}-x_{0}\right) S^{+}\left( x,x^{\prime }\right) \,,  \notag \\
&&S^{-}\left( x,x^{\prime }\right) =i\sum_{n,m}{}^{+}\psi _{n}\left(
x\right) G\left( \left. _{+}\right| ^{+}\right) _{nm}^{-1}\,_{+}\bar{\psi}%
_{m}\left( x^{\prime }\right) \,,  \notag \\
&&S^{+}\left( x,x^{\prime }\right) =i\sum_{n,m}\,_{-}\psi _{n}\left(
x\right) \left[ G\left( \left. _{-}\right| ^{-}\right) ^{-1}\right]
_{nm}^{\dagger }\,^{-}\bar{\psi}_{m}\left( x^{\prime }\right) \,,
\label{emt19}
\end{eqnarray}
see \cite{Gitma77,FGS91}. Then, the difference $S^{p}(x,x^{\prime })=\
S_{in}^{c}(x,x^{\prime })-S^{c}(x,x^{\prime })$ has the form 
\begin{equation}
\ S^{p}(x,x^{\prime })=i\sum_{nm}\,_{-}{\psi }_{n}(x)\,\left[
G(_{+}|^{-})G(_{-}|^{-})^{-1}\right] _{nm}^{\dagger }{_{+}\bar{\psi}}%
_{m}(x^{\prime })\,.  \label{emt21}
\end{equation}
This function vanishes in the case of a stable vacuum, since it contains the
coefficients $G(_{+}|^{-})$ related to the mean number of created particles.

Thus, the mean EMT (\ref{emt22}) can be presented in such a form that
contributions of different kinds are explicitly separated, namely, 
\begin{equation}
\langle T_{\mu \nu }\left( t\right) \rangle =\langle T_{\mu \nu }\left(
t\right) \rangle ^{0}+\langle T_{\mu \nu }\left( t\right) \rangle ^{\theta
},\;\;\langle T_{\mu \nu }\left( t\right) \rangle ^{0}=\func{Re}\langle
T_{\mu \nu }\left( t\right) \rangle ^{c}+\func{Re}\langle T_{\mu \nu }\left(
t\right) \rangle ^{p}.  \label{emt24.1}
\end{equation}
The quantities $\langle T_{\mu \nu }\left( t\right) \rangle
_{\;\;\;\;\;}^{c,p,\theta }$ are defined as 
\begin{eqnarray}
&&\,\langle T_{\mu \nu }\left( t\right) \rangle _{\;\;\;\;\;}^{c,p,\theta
}=i\left. \mathrm{tr}\left[ A_{\mu \nu }S^{c,p,\theta }(x,x^{\prime })\right]
\right| _{x=x^{\prime }}\,,  \notag \\
&&A_{\mu \nu }=1/4\left[ \gamma _{\mu }\left( P_{\nu }+P\mathcal{^{\prime }}%
_{\nu }^{\ast }\right) +\gamma _{\nu }\left( P_{\mu }+P\mathcal{^{\prime }}%
_{\mu }^{\ast }\right) \right] \,,  \label{emt24.2}
\end{eqnarray}
where $\mathrm{tr}\left[ \cdots \right] $ is the trace in the space of $%
4\times 4$ matrices.

Calculating components $\func{Re}\,\langle T_{\mu \nu }\left( t\right)
\rangle ^{c}$ in the case of the $T$-constant field being subject to
condition{\large \ }(\ref{unistab}), we find that they are connected with
the real part of the Heisenberg--Euler Lagrangian $\mathcal{L}$, namely, 
\begin{eqnarray}
&\,&\func{Re}\langle T_{00}(t)\rangle ^{c}=-\func{Re}\langle
T_{33}(t)\rangle ^{c}=E\frac{\partial \func{Re}\mathcal{L}}{\partial E}-%
\func{Re}\mathcal{L}\,,  \notag \\
&&\func{Re}\,\langle T_{11}(t)\rangle ^{c}=\func{Re}\langle T_{22}(t)\rangle
^{c}=\func{Re}\mathcal{L}-B\frac{\partial \func{Re}\mathcal{L}}{\partial B}%
\,.  \label{HEL}
\end{eqnarray}

These components describe the contribution due to the vacuum polarization.
Their field-dependent parts are finite after a standard renormalization and
do exist for arbitrary quasiconstant electric field. They are local, i.e.,
they depend on $t$, but do not depend on the history of the process. The
contributions $\func{Re}\langle T_{\mu \nu }(t)\rangle ^{p}$ arise due to
vacuum instability. The quantity $<T_{\mu \nu }\left( t\right) >^{\theta }$
presents the contribution due to the existence of the initial thermal
distribution.

III. We have calculated the functions $S^{p}$ and $S^{\theta }$ explicitly;
see the details in \cite{GavG07}. Using these functions, we have obtained a
representation for the mean EMT $\langle T_{\mu \nu }(t)\rangle $. The
expression consists of three terms:

\begin{equation}
\,\langle T_{\mu \nu }(t)\rangle =\func{Re}\langle T_{\mu \nu }(t)\rangle
^{c}+\mathrm{\func{Re}}\langle T_{\mu \nu }\left( t\right) \rangle _{\theta
}^{c}+\tau _{\mu \nu }^{p}\left( t\right) \,.  \label{t2}
\end{equation}
As mentioned above, the term $\func{Re}\langle T_{\mu \nu }(t)\rangle ^{c}$
is the contribution due to vacuum polarization. The term $\mathrm{\func{Re}}%
\langle T_{\mu \nu }\left( t\right) \rangle _{\theta }^{c}$ describes the
contribution due to the work of the external field on the particles at the
initial state. The term $\tau _{\mu \nu }^{p}\left( t\right) $ describes the
contribution due to particle creation from vacuum. The two latter
contributions depend on the time interval $t-t_{1}$, and therefore they are
global quantities. The term $\tau _{\mu \nu }^{p}\left( t\right) $ includes
the factor $\exp \left\{ -\pi M^{2}/\left| qE\right| \right\} $. This factor
is exponentially small for a weak electric field, $M^{2}/\left| qE\right|
\gg 1$, whereas the term is not small as long as the electric field strength
approaches the critical value $E_{c}$. On the other hand, the term $\mathrm{%
\func{Re}}\langle T_{\mu \nu }\left( t\right) \rangle _{\theta }^{c}$, as
well as $\func{Re}\langle T_{\mu \nu }(t)\rangle ^{c}$, does exist in
arbitrary electric field. When the $T$-constant electric field turns off (at 
$t>t_{2}$), the local contribution to $\mathrm{\func{Re}}\langle T_{\mu \nu
}(t)\rangle ^{c}$ made by the electric field becomes equal to zero; whereas
the global contributions, given by $\mathrm{\func{Re}}\langle T_{\mu \nu
}\left( t\right) \rangle _{\theta }^{c}$ and $\tau _{\mu \nu }^{p}\left(
t\right) $, do not vanish and retain their values at any $t>t_{2}$.

In the case of a strong electric field, we separate the increasing terms
related to a large interval $t-t_{1}$ and call them the ``leading
contributions''. They appear due to particle creation. However, the
quantities $\mathrm{\func{Re}}\langle T_{\mu \nu }\left( t\right) \rangle
_{\theta }^{c}$ are computed for any $t$, $0\leq t-t_{1}\leq T$. All these
contributions are investigated in detail at different regimes and limits of
weak and strong fields, as well as at low and high temperatures. For all
these limiting cases, we have obtained the leading contributions, which are
given by elementary functions of the fundamental dimensionless parameters.
These results are presented below.

In a weak electric field, $\left| qE\right| /M^{2}\ll 1$, at low
temperatures, $M\beta \gg 1,$ for $\left| \mu ^{(\zeta )}\right| \ll M$, the
nonzero expressions for $\func{Re}\langle T_{\mu \nu }\left( t\right)
\rangle _{\theta }^{c}$ read as follows:

a) when the increment of kinetic momentum is small, $\left| qE\right| \left(
t-t_{1}\right) /M\ll 1$, we have 
\begin{equation}
\func{Re}\langle T_{\mu \nu }\left( t\right) \rangle _{\theta }^{c}=\bar{T}%
_{\mu \nu }^{0}+\Delta T_{\mu \nu },  \label{emt97}
\end{equation}
and 
\begin{eqnarray}
&&\Delta T_{11}=\Delta T_{22}=-C\frac{2}{M\beta }\left[ 1+\frac{M\left(
t-t_{1}\right) ^{2}}{\beta }\right] \left[ 1+O\left( \frac{1}{M\beta }%
\right) \right] ,  \notag \\
&&\Delta T_{33}=C\left[ -\frac{3}{2M\beta }+4\frac{M\left( t-t_{1}\right)
^{2}}{\beta }\right] \left[ 1+O\left( \frac{1}{M\beta }\right) \right] , 
\notag \\
&&\Delta T_{00}=C\left[ -\frac{235}{256}+2\frac{M\left( t-t_{1}\right) ^{2}}{%
\beta }\right] \left[ 1+O\left( \frac{1}{M\beta }\right) \right] ,  \notag \\
&&C=\frac{\left( qE\right) ^{2}}{2\pi ^{2}}\left( \frac{\pi }{2M\beta }%
\right) ^{1/2}e^{-M\beta }\,,  \label{emt96}
\end{eqnarray}
where we have explicitly separated the part $\bar{T}_{\mu \nu }^{0}=\left.
\langle T_{\mu \nu }\left( t\right) \rangle _{\theta }^{c}\right| _{E=0}$,
being independent of the electric field;

b) when the increment of kinetic momentum is large, $\left| qE\right| \left(
t-t_{1}\right) /M\gg 1$, we have 
\begin{eqnarray}
\func{Re}\langle T_{11}\left( t\right) \rangle _{\theta }^{c} &=&\func{Re}%
\langle T_{22}\left( t\right) \rangle _{\theta }^{c}=\frac{M}{\left|
qE\right| \left( t-t_{1}\right) }\bar{T}_{11}^{0}\,,  \notag \\
\func{Re}\langle T_{00}\left( t\right) \rangle _{\theta }^{c} &=&\func{Re}%
\langle T_{33}\left( t\right) \rangle _{\theta }^{c}=\left| qE\right| \left(
t-t_{1}\right) \sum_{\zeta =\pm }n^{(\zeta )}\,,  \label{103}
\end{eqnarray}
where\ $n^{(\zeta )}$ is the initial particle density, and $\bar{T}_{11}^{0}$
is the initial value of $\func{Re}\langle T_{11}\left( t\right) \rangle
_{\theta }^{c}$ at $t<t_{1}$.

At high temperatures, $M\beta \ll 1\,,\;\;\sqrt{\left| qB\right| }\beta \ll
1\,,\;\;\sqrt{\left| qE\right| }\beta \ll 1\,$, and in case the increment of
kinetic momentum is small, $\left| qE\right| \left( t-t_{1}\right) \beta \ll
1$, the nonzero expressions for $\func{Re}\langle T_{\mu \nu }\left(
t\right) \rangle _{\theta }^{c}$ have the form (\ref{emt97}), while the
terms $\Delta T_{\mu \nu }$ depending on the electric field have the form 
\begin{eqnarray}
&&\Delta T_{11}=\Delta T_{22}=\frac{\left( qE\right) ^{2}}{12\pi ^{2}}\left[
-\frac{1}{2}\ln \left( M\beta \right) +O\left( 1\right) -\frac{29\pi ^{2}}{15%
}\frac{\left( t-t_{1}\right) ^{2}}{\beta ^{2}}\right] \,,  \notag \\
&&\Delta T_{33}=\frac{\left( qE\right) ^{2}}{12\pi ^{2}}\left[ -\ln \left(
M\beta \right) +O\left( 1\right) -\frac{7\pi ^{2}}{15}\frac{\left(
t-t_{1}\right) ^{2}}{\beta ^{2}}\right] \,,  \notag \\
&&\Delta T_{00}=\frac{\left( qE\right) ^{2}}{12\pi ^{2}}\left[ -2\ln \left(
M\beta \right) +O\left( 1\right) -\frac{13\pi ^{2}}{3}\frac{\left(
t-t_{1}\right) ^{2}}{\beta ^{2}}\right] \,.  \label{emt116}
\end{eqnarray}
In case the increment of kinetic momentum is large, $\left| qE\right| \left(
t-t_{1}\right) \beta \gg 1$, the nonzero expressions for $\func{Re}\langle
T_{\mu \nu }\left( t\right) \rangle _{\theta }^{c}$ decrease exponentially: 
\begin{equation}
\func{Re}\langle T_{\mu \mu }\left( t\right) \rangle _{\theta }^{c}\sim \exp %
\left[ -\left| qE\right| \left( t-t_{1}\right) \beta \right] .
\label{emt120}
\end{equation}

Before the $T$-constant field turns on, the system under consideration is at
thermal equilibrium. The system is described by the thermodynamic potential $%
\Omega =-\theta \ln Z$, as well as by the temperature- and field-dependent
renormalized effective Lagrangian $\mathcal{L}^{\theta }=-\Omega /V$, where
the partition function $Z$ is defined by (\ref{emt1}). The quantity $\left.
\Omega \right| _{B=0}$ is well-known from textbooks. In case $\mu
^{(+)}=-\mu ^{(-)}=\mu $, the expression for $\mathcal{L}^{\theta }$ was
presented in \cite{CanD96,ElmPS93}, and at $\mu =0$ was obtained in \cite
{Dit79}. It turns out that the nonzero components of $\bar{T}_{\mu \nu }^{0}$
in (\ref{emt97}) can be deduced from the effective Lagrangian $\mathcal{L}%
^{\theta }$ as follows: 
\begin{eqnarray}
\bar{T}_{00}^{0} &=&-\mathcal{L}^{\theta }-\beta \frac{\partial \mathcal{L}%
^{\theta }}{\partial \beta }+\sum_{\zeta =\pm }\mu ^{(\zeta )}\frac{\partial 
\mathcal{L}^{\theta }}{\partial \mu ^{(\zeta )}}\,,  \notag \\
\bar{T}_{11}^{0} &=&\bar{T}_{22}^{0}=\mathcal{L}^{\theta }-B\frac{\partial 
\mathcal{L}^{\theta }}{\partial B}\,,\;\;\bar{T}_{33}^{0}=\mathcal{L}%
^{\theta }\,.  \label{emt84a}
\end{eqnarray}
Thus, our results for $\langle T_{\mu \nu }\left( t\right) \rangle _{\theta
}^{c}$ are in agreement with the results obtained previously in the case $%
E=0 $.

For $t\geq t_{2}$, particle production is absent, and the quantity $\tau
_{\mu \nu }^{p}\left( t_{2}\right) $ presents the mean EMT of created pairs.
At low temperatures, $\beta \left| \left( \varepsilon _{\mathbf{p}}^{(\zeta
)}-\mu ^{(\zeta )}\right) \right| \gg 1$, under the assumption $\varepsilon
_{\mathbf{p}}^{(\zeta )}>\left| \mu ^{(\zeta )}\right| $, the leading
contributions in $\tau _{\mu \nu }^{p}\left( t_{2}\right) $ coincide with
the vacuum contributions: 
\begin{eqnarray}
\tau _{\mu \nu }^{p}\left( t_{2}\right) &=&\tau _{\mu \nu }^{cr}\left(
t_{2}\right) ,\;\;\tau _{00}^{cr}\left( t_{2}\right) =\tau _{33}^{cr}\left(
t_{2}\right) =\left| qE\right| Tn^{cr},  \notag \\
\tau _{11}^{cr}\left( t_{2}\right) &=&\tau _{22}^{cr}\left( t_{2}\right)
\sim \ln \left( \sqrt{\left| qE\right| }T\right) .  \label{t3}
\end{eqnarray}
Here, $n^{cr}$ is the total number-density of pairs created by the $T$%
-constant electric field, 
\begin{equation*}
n^{cr}=\frac{q^{2}}{4\pi ^{2}}EBT\,\coth (\pi B/E)\exp \left( -\pi \frac{%
M^{2}}{\left| qE\right| }\right) .
\end{equation*}
At high temperatures, $\beta \left| qE\right| T\ll 1$, we have 
\begin{eqnarray}
&&\tau _{00}^{p}\left( t_{2}\right) =\tau _{33}^{p}\left( t_{2}\right) =%
\frac{1}{6}\beta \left| qE\right| T\tau _{00}^{cr}\left( t_{2}\right) \,, 
\notag \\
&&\tau _{11}^{p}\left( t_{2}\right) =\tau _{22}^{p}\left( t_{2}\right) =%
\frac{1}{2}\beta \left| qE\right| T\tau _{11}^{cr}\left( t_{2}\right) .
\label{emt76}
\end{eqnarray}

IV. The action of electric field manifests itself differently for different
components of the EMT $\langle T_{\mu \nu }(t)\rangle $. One can see that in
a strong electric field, or in a field acting during a long time interval,
the leading terms for the energy density,\ $\mathrm{\func{Re}}\langle
T_{00}\left( t\right) \rangle _{\theta }^{c}+\tau _{00}^{p}\left( t\right) $%
, and those for the pressure{\large \ }along the direction of the electric
field, $\mathrm{\func{Re}}\langle T_{33}\left( t\right) \rangle _{\theta
}^{c}+\tau _{33}^{p}\left( t\right) $, are identical. However, for vacuum
polarization, these terms have opposite signs: $\func{Re}\langle
T_{00}(t)\rangle ^{c}=-\func{Re}\langle T_{33}(t)\rangle ^{c}$. This is due
to a difference between the equation of state for the relativistic fermions
and the equation of state for the electromagnetic field. The dependence of
the electric field and its duration for the transversal component $\mathrm{%
\func{Re}}\langle T_{11}\left( t\right) \rangle _{\theta }^{c}+\tau
_{11}^{p}\left( t\right) $ is quite different from the dependence of the
longitudinal component $\mathrm{\func{Re}}\langle T_{33}\left( t\right)
\rangle _{\theta }^{c}+\tau _{33}^{p}\left( t\right) $. This implies that
one cannot define a universal functional of the electric field whose
variations should produce all the components of EMT. This fact holds true
equally for vacuum and non-vacuum initial states. We believe that for
particle-creating backgrounds one cannot obtain any generalization of the
Heisenberg--Euler Lagrangian for the mean values. This fact is connected
with the existence of non-local contributions to the EMT. This may explain
the failure of numerous attempts at considering one-loop effects on the
basis of different variants of such a generalization. Our results also
demonstate that in particle-creating backgrouns any generalization of the
Heisenberg--Euler Lagrangian to the case of finite temperatures is
problematic.

One can neglect the back-reaction of created pairs in strong electric field
only in the case $\tau _{00}^{p}\left( t_{2}\right) \ll $ $E^{2}/8\pi $,
which implies a restriction on the strength and duration of the electric
field. This restriction has the same form $\left| qE\right| T^{2}\ll \frac{%
\pi ^{2}}{2q^{2}}$ for both the initial vacuum state and the low-temperature
initial thermal state. On the other hand, there exists another restriction, $%
1\ll \left| qE\right| T^{2}$ (see (\ref{unistab})), which allows one to
disregard the details of switching the electric field on and off. Collecting
these two restrictions, we obtain a range of the dimensionless parameter $%
\left| qE\right| T^{2}$ for which QED with a strong constant electric field
is consistent: 
\begin{equation*}
1\ll \left| qE\right| T^{2}\ll \pi ^{2}/2q^{2}\,.
\end{equation*}

Similar restrictions can take place when the initial thermal equilibrium is
examined at sufficiently high temperatures, $\beta \left| qE\right| T\ll 1$.
In this case, we have two inequalities, $\beta \left| qE\right| ^{2}T^{3}\ll 
\frac{3\pi ^{2}}{q^{2}}\,$and $1\ll \left| qE\right| T^{2}$, which imply 
\begin{equation*}
1\ll \left| qE\right| T^{2}\ll \frac{3\pi ^{2}}{q^{2}\beta \left| qE\right| T%
}\,.
\end{equation*}
We can see that the upper restriction for $\left| qE\right| T^{2}$ is weaker
than it is in the low-temperature case.

\subparagraph{\protect\large Acknowledgement}

S.P.G. thanks FAPESP for support and Universidade de São Paulo for
hospitality. D.M.G. acknowledges FAPESP and CNPq for permanent support.

\end{document}